\documentclass[useAMS,usenatbib,usegraphicx]{mn2e}

\newcommand{\co}{CO}

\newcommand{\thco}{$^{13}$CO}

\newcommand\aj{{AJ}}%
\newcommand\apj{{ApJ}}%
\newcommand\apjl{{ApJ}}%
\newcommand\apjs{{ApJS}}%
\newcommand\aap{{A\&A}}%
\newcommand\mnras{{MNRAS}}%
\begin{document}

\title[HCN in M33]{Minimal HCN emission from Molecular Clouds in M33}

\author[Rosolowsky, Pineda \& Gao]{Erik Rosolowsky,$^1$, Jaime E.
  Pineda,$^{2,3}$, Yu Gao$^4$\\ 
$^1$ University of British Columbia Okanagan, 3333 University Way,
Kelowna BC V1V 1V7 Canada\\
$^2$ Harvard-Smithsonian Center for Astrophysics, 60 Garden St., MS-10, Cambridge, MA 02138, USA\\
$^3$ Jodrell Bank Centre for Astrophysics, School of Physics and Astronomy, University of Manchester, Oxford Road, Manchester, M13 9PL, UK\\
$^4$ Purple Mountain Observatory, Chinese Academy of Sciences (CAS), 2 West Beijing Road, Nanjing 210008, China}
\maketitle
\begin{abstract}
Since HCN emission has been shown to be a linear tracer of ongoing
star formation activity, we have searched for HCN $(J=1\to 0)$
emission from known GMCs in the nearby galaxy M33.  No significant HCN
emission has been found along any of the lines of sight.  We find two
lines of sight where CO-to-HCN integrated intensity ratios up to
$280$, nearly a factor of 6 above what is found in comparable regions
of the Milky Way.  Star formation tracers suggest that the HCN-to-star
formation rate ratio ($L_{\mathrm{HCN}}/\dot{M}_{\star}$) is a factor
of six lower than what is observed in the Milky Way (on average) and
local extragalactic systems.  Simple chemical models accounting for
the sub-solar N/O ratio suggest that depletion cannot account for the
high CO-to-HCN ratios.  Given HCN formation requires high extinction
($A_V>4$), low metallicity may yield reduced dust shielding and thus a
high CO/HCN ratio.  The turbulence and structure of GMCs in M33 are
comparable to those found in other systems, so the differences are
unlikely to result from different GMC properties.  Since lower
CO-to-HCN ratios are associated with the highest rates of star
formation, we attribute the deficits in part to evolutionary effects
within GMCs.
\end{abstract}
\section{Introduction}

All local star formation is invariably associated with molecular gas,
but recent studies have clarified that star formation is associated
exclusively with {\em dense} ($n> 10^5 \mbox{ cm}^{-3}$) molecular gas
from the scales of cores \citep{lada-dense,wu05-hcn,wu10} to galaxies
\citep{gs04,gao07}.  Most of the gas typically observed in
studies of molecular clouds on large scales ($>10$ pc) is found at
lower densities ($n\sim 10^2 \mbox{ cm}^{-3}$), but dense gas studies
suggest that the low-density gas is not directly involved in the star
formation process.  This clarification is particularly important for
studies of the galaxy-scale star formation since the molecular gas
content of a external galaxy is usually traced with emission from
low-$J$ transitions of $^{12}$CO, which are excited in the low density
molecular gas which traces the bulk of material found in molecular
clouds \citep[e.g.,][]{pineda-abund}.

Relatively few large-scale studies of the dense molecular gas content
of galaxies have been conducted. The majority of mass in the dense gas
is found as molecular hydrogen, which emits no radiation at the
densities and temperatures typical of star forming clouds.
Observations rely on emission from dense gas tracers, which are
excited to emission only at the high densities.  Typical tracers
include line emission from HCN, N$_2$H$^{+}$, NH$_3$ and HCO$^+$, all
of which have critical densities of order $10^5\mbox{ cm}^{-3}$.
While the relative merits of these tracers have been compared
\citep{shirley08}, all seem to produce similar results: the dense gas
that they trace is linearly related to the star formation rate
\begin{equation}
\frac{\Sigma_{\mathrm{SFR}}}{M_{\odot}~\mathrm{yr^{-1}~pc^{-2}}}
\propto
\left(\frac{\Sigma_{\mathrm{dense}}}{M_{\odot}~\mathrm{pc^{-2}}}\right)^{1.0}.
\end{equation}
These results even seem to hold for high-redshift systems
\citep{gao07}. 

Even though low-density molecular gas does not participate directly in
star formation, there has been extensive work demonstrating that
low-density gas is indeed related to star formation on a galactic
scale.  While originally demonstrated for galaxies as a whole
\citep{k98}, recent work has shown the relationship holds for
molecular gas on scales of $> 500$ pc
\citep{wb02,m33-fcrao,k07-short,bigiel08}.  These studies find
$\Sigma_{\mathrm{SFR}} \propto \Sigma_{\mathrm{H2}}^{\alpha}$, where
$\alpha=1.0$ to $1.4$ depending on analysis methods used.  The apparent
difference in star formation relationships for low- and high-density
molecular gas can likely be attributed to changing dense-gas fractions
in a galaxy's ISM \citep{gs04}.  Some authors argue that the index of
the star formation law can be predicted from the average density of
the medium compared to the critical density of tracer in question
\citep{krum-ksslope,narayanan-ks}, with additional considerations
regarding excitation of higher levels. Even with these considerations,
the $(1\to 0)$ line of HCN should provide a linear tracer of dense gas
mass and an index of unity on the star formation law.  Thus, a
compelling model emerges where the star formation process in a galaxy
can be decoupled into two steps: the formation of dense gas within a
molecular cloud and the subsequent formation of stars in that
high-density gas {\it at a constant rate} \citep[or a constant
  fraction per free-fall time,][]{krum-ksslope}.  The apparent changes
in the star formation efficiency
($\Sigma_{\mathrm{H2}}/\Sigma_{\mathrm{SFR}}$) can then be attributed
to change in the molecular cloud properties in galaxies, specifically
the fraction of the cloud found at high density.

Comparing the CO and HCN emission from individual molecular clouds can
test whether there is significant change in the fraction of dense gas
within individual molecular clouds.  Some studies have been undertaken
for Milky Way molecular clouds \citep[e.g.,][]{hcn-MW}, suggesting
that dense gas fraction of molecular gas does indeed change as a
function of distance from the centre of the Galaxy. The study of
\citet{m31-hcn} examined the CO-to-HCN intensity ratio across the disk
of the massive galaxy M31, finding a radial decline in the ratio with
a maximum ratio of $I_{\mathrm{CO}}/I_{\mathrm{HCN}}=125$.  It remains
unclear based on these results the degree to which the changes in the
CO-to-HCN ratio result from changes in molecular clouds properties or
whether the relationship between the tracers and the underlying H$_2$
is changing.  

The nearby galaxy M33 has been noted to exhibit a particularly high
star formation efficiency \citep{m33-fcrao, deepm33, gardan-m33} based
on observations of CO emission.  In the context of the above star
formation model, this would imply that the dense gas fractions of the
molecular clouds in M33 would be higher than that of the typical
galaxy.  Other circumstances may explain the high star formation
efficiency of the galaxy since M33 is distinct from M31 and the Milky
Way in two ways.  First, it has a significantly lower metallicty
outside the central region \citep[$12+\log(\mbox{O/H})
  =8.4$][]{magrini07,m33-grad}; and second, it has a significantly
lower mass ($M_{\mathrm{M33}}\sim M_{\mathrm{MW}}/10$).  Given its
proximity and previous study it is possible to assess how tracers of
high- and low-density molecular gas relate on the scale on individual
giant molecular clouds.  To this end, we have undertaken observations
of CO and HCN toward giant molecular clouds (GMCs) in M33 using the
IRAM 30-m telescope. 

\section{Molecular Line Observations}
We observed the positions of four Giant Molecular Clouds (GMCs) in M33
using the IRAM 30-m telescope located in Pico Veleta, Spain.  The
observations were carried out from 16th to 20th of July 2007.
HCN~($1\to 0$) and \thco~($2\to 1$) were observed simultaneously with
the VESPA back-end in the $80$~kHz resolution mode.  The same
configuration was used to observe $^{12}$\co~($1\to 0$) and
$^{12}$\co~($2\to 1$) simultaneously. All the observations were taken
in position-switching mode.  Weather conditions were good for summer
observations and pointing observations imply a pointing accuracy of
$<5''$.  Of note, this accuracy is small compared to the 89 GHz beam
(HCN($1\to 0$)) but is more significant when compared to the 230 GHz
beam ($^{12}$CO($2\to 1$)).  Opacity and system temperature for each of the
transitions is given in Table \ref{tab:obs} and integration time
($t_{int}$) and per-channel rms ($\sigma$) for the HCN/$^{13}$CO
configuration is given in Table \ref{gmctable}.

\begin{figure}
\includegraphics[width=84mm]{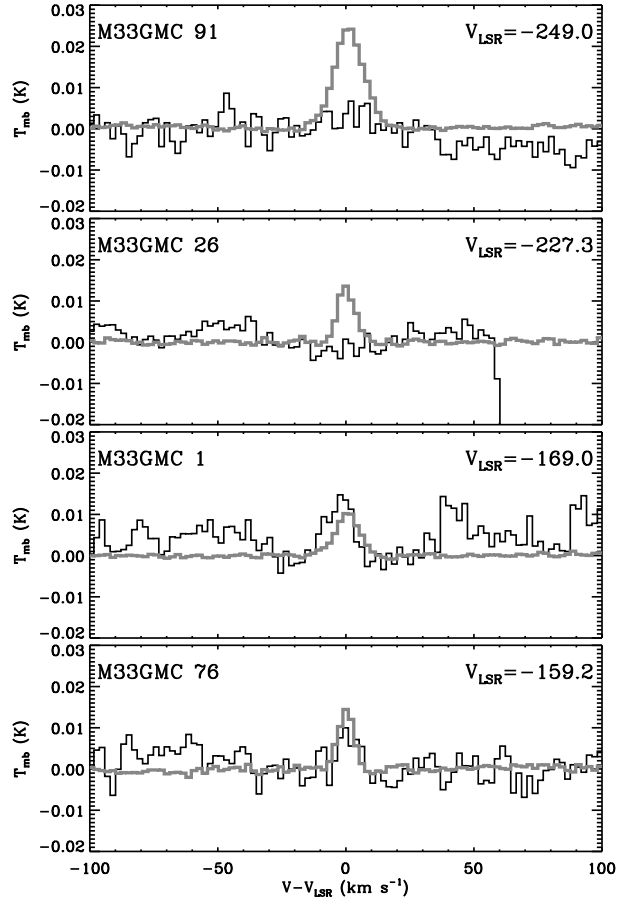}
\caption{\label{COfig} IRAM 30-m observations of HCN ($1\to 0$) and
  $^{12}$CO ($1\to 0$) shown as black and gray lines respectively.
  The CO spectra have been scaled down by a factor of 100 for ease of
  comparison with the HCN data.  The spectra have been Hanning
  smoothed and decimated to 2 km s$^{-1}$ resolution.}
\end{figure}

Conversion from the measured $T_A^*$ to $T_{\mathrm{mb}}$ is
carried out as $T_{\mathrm{mb}}=T_A^* \,F_{eff}/B_{eff}$.  The forward
($F_{eff}$) and beam ($B_{eff}$) efficiencies were determined by
measurements made by observatory staff.  A summary of the
observation's parameters is presented in Table~\ref{tab:obs} and the
spectra are shown in Figure \ref{COfig}.  The data were reduced using
the CLASS90 software.

\begin{table}
\caption{Observational parameters \label{tab:obs}}
\begin{tabular}{@{}lcccccc}
\hline
Line & 
Freq. & 
$F_{eff}$ &
$B_{eff}$ & 
$\theta_{\mathrm{FWHM}}$ & $\langle T_{sys}\rangle$ & $\langle \tau\rangle$\\
& (GHz) & & & (arcsec) & (K) & \\
\hline
$^{12}{\rm CO} (1\rightarrow0)$ & 115.2712 & 0.95 &  0.75 & 22 & 250 &
0.28\\
$^{12}{\rm CO} (2\rightarrow1)$ & 230.538 & 0.91 &  0.52 & 11 & 250 & 0.11\\
$^{13}{\rm CO} (2\rightarrow1)$ & 220.3986 & 0.91 &  0.57 & 12 & 250 &
0.11\\
${\rm HCN} (1\rightarrow0)$ & 88.63160 & 0.95 &  0.78 & 28& 110 & 0.05 \\
\hline
\end{tabular}
\end{table}

\begin{figure}
\includegraphics[width=84mm]{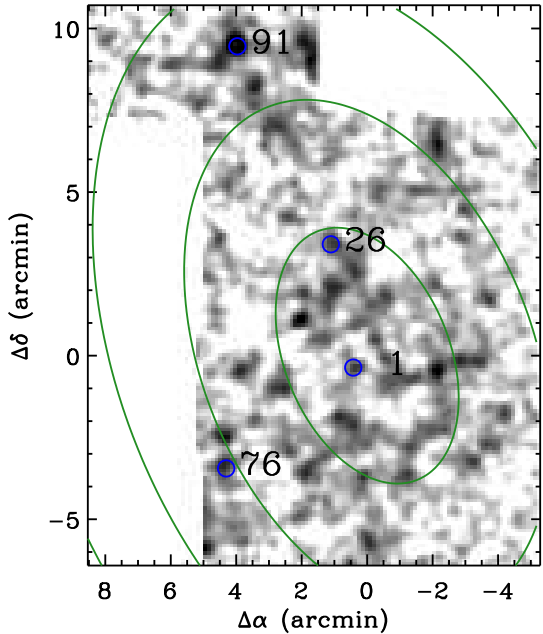}
\caption{\label{comap} Locations of the four IRAM lines of sight
  overlaid on an integrated intensity of CO ($1\to 0$) emission.
  Three contours of constant galactocentric radius are indicated with
  ellipses representing $R_{\mathrm{kpc}}=1,2$ and 3 kpc.}
\end{figure}

We selected four targets from the GMC catalog of \citet{deepm33} based
on strong CO emission with a range of star formation properties.  We
summarize the observations of the GMCs in Table \ref{gmctable} and
their locations are shown in Figure \ref{comap}.  We did not detect
HCN ($1\to 0$) emission in at least two of the four lines of sight.
M33GMC 1, the pointing closest to the center of the galaxy, shows a
modest detection at a low level, as does M33GMC 76.  For each
line-of-sight, we measure the integrated intensity $I_{\mathrm{HCN}} =
\sum T_{\mathrm{mb}} \delta V$ where $\delta V$ is the channel width,
where the sum is carried out over the expected velocity range of the HCN
line assuming it has the same intrinsic line width as the CO line but
accounting for hyperfine structure.  
We establish a 3$\sigma$ upper limit as
\begin{equation}
I_{\mathrm{HCN}}< 3\sigma_{\mathrm{HCN}} \Delta V \left[(1-N/N_{tot})\right]^{-0.5},
\end{equation}
where $\sigma_{\mathrm{HCN}}$ is the rms noise fluctuation in the
spectrum, $\Delta V$ is the velocity width of the HCN line, $N$ is the
number of channels across $\Delta V$ and $N_{tot}$ is the number of
channels in the entire spectrum \citep{matthews-gao}. 

We further process these observations to place them on a common
standard and supplement the results with a significant amount of
ancillary data, the results of which are given in Table
\ref{gmctable}.  The integrated intensities are given as $I_{10}$,
$I_{21}$ and $I_{13}$ for the $^{12}$CO($1\to 0$), $^{12}$CO($2\to
1$), and $^{13}$CO($1\to 0$) lines respectively.  For comparison to
other work, molecular line intensities are scaled to luminosities for
the lines by multiplying by the projected area of the beam for that
tracer.

For a rigourous comparison of line intensities, all data must be sampled with
the same resolution.  To properly estimate the CO-to-HCN ratio, we convolved
the fully sampled, high resolution $^{12}$CO$(1\to 0$) map of \citet{deepm33}
to the $28''$ resolution of the IRAM beam at 89 GHz and sampled the map at the
observed positions.  These values are reported as $I_{10,\mathrm{R07}}$ and
$I_{10,\mathrm{convol}}$ in Table \ref{gmctable} for pre- and post-convolution
respectively. We checked the relative calibrations of the map by
convolving the same map to the $22''$ resolution of the IRAM
$^{12}\mathrm{CO}~(1\to 0)$ observations and comparing the observed and
predicted integrated intensities finding them consistent to the 10\% level
expected from the amplitude calibrations of the two maps.

We supplement the molecular line data with infrared and optical band
observations which are indicators of the star formation rate.  To
derive the star formation rate, we use combined H$\alpha$+24~$\mu$m
emission derived from data from the Local Group Survey
\citep{massey-halpha} and Spitzer/MIPS observations of the galaxy
\citep{spitzer-mips} respectively.  A continuum subtracted
H$\alpha$ image was generated by subtracting off an optimally scaled
$R$ band image from the H$\alpha$ filter and scaling to photometric
units based on \citet{massey-halpha}.  The MIPS data were from the
calibrated images of \citet{hinz-m33} and \citet{tab-m33}.  Each image
was convolved to the $28''$ IRAM beam and sampled at the locations of
the IRAM pointings.  The local star formation was established using
the prescription of \citet{k07}:
\begin{equation}
\dot{M}_{\star}~(M_{\odot}~\mbox{yr}^{-1}) = 7.9\times
10^{-42}\left[L_{\mathrm{H}\alpha}+0.039~
\nu_{24}L_{\nu}(24)\right]
\end{equation}
where luminosities are measured in erg~s$^{-1}$ and refer to the H$\alpha$
line and the Spitzer 24-micron band respectively.  

We also estimate the total infrared luminosity using all three MIPS
bands following \citet{dale-helou}.
\begin{equation}
L_{\mathrm{TIR}} = w_1\nu L_\nu(24~\mu m)+ 
w_2\nu L_\nu(70~\mu m)+ w_3 \nu L_\nu(160~\mu m),
\end{equation}
where $w_1=1.559$, $w_2=0.7686$ and $w_3=1.6381$.  The MIPS 160~$\mu$m band
image is sampled without convolution since the resolution is larger than the
HCN: $40''$ vs. $28''$, resulting in an additional uncertainty in $L_{\mathrm{TIR}}$ of
15\%.  We also estimate the local pressure in the galaxy using the estimator
presented by \citet{pressure2} and we report these values of $P/k$ in Table
\ref{gmctable}.  The total mass of the observed GMC as determined from the
high-resolution map (R07) is given as $M_{\mathrm{CO}}$.  The ``Comparisons''
section of the table highlights the most important results as discussed in
\S\ref{analysis} below.

\begin{table*}
\begin{minipage}{126mm}
\caption{M33 GMCs observed \label{gmctable}}
\begin{tabular}{@{}lrrrr}
\hline
Property & M33GMC 91 & M33GMC 26 & M33GMC 1 &M33GMC 76\\
\hline
$\alpha_{2000}$ & 01~34~09.2 & 01~33~55.8 & 01~33~52.4 &  01~34~10.7\\
$\delta_{2000}$ & +30~49~06 & +30~43~02 & +30~39~18 & +30~36~15 \\
$V_{LSR}~(\mbox{km s}^{-1})$ & $-249.0$ & $-227.3$ & $-169.0$ & $-159.2$ \\
$t_{int}~(\mbox{min})$ &246  &217 & 244 &213 \\
$\sigma~(\mbox{mK})$ & 7.1 & 3.1 & 5.6 & 5.5 \\
$I_{10} ~(\mbox{K km s}^{-1})$ & 21.6 & 6.9 & 7.2 & 6.5 \\
$I_{21} ~(\mbox{K km s}^{-1})$ & 25.2 & 7.0 & 15.5 & 9.6 \\
$I_{13} ~(\mbox{K km s}^{-1})$ & 1.4 & 0.3 & 1.8 & 1.6 \\
$I_{\mathrm{HCN}} ~(\mbox{K km s}^{-1})$ & $< 0.053$ & $< 0.027$
&$\sim 0.12$ & $\sim 0.059$ \\
$L_{10}~(10^3\mbox{ K km s}^{-1}\mbox{ pc}^{2})$$^a$ & 150 & 53 & 47 & 59
\\
$L_{\mathrm{HCN}}~(10^3\mbox{ K km s}^{-1}\mbox{
  pc}^{2})$$^a$ & $<0.75$ & $<0.40$ &$\sim 1.90$ & $\sim 0.94$ \\
\hline
Ancillary Data \\
\hline
$I_{10,\mathrm{R07}}~(\mbox{K km s}^{-1})$ & 29.0 & 12 & 11 & 11 \\
$I_{10,\mathrm{convol}}~(\mbox{K km s}^{-1})$ & 13.1 & 5.0 & 4.4 & 5.2 \\
$L_{\mathrm{H}\alpha}~(10^3~L_{\odot})$$^a$ &  1.4 & 2.9 & 34 & 10 \\
$\nu_{24}L_{\nu}(24)~(10^3~L_{\odot})$$^a$ & 61 & 78 & 470 & 250 \\
$L_{\mathrm{TIR}}~(10^6~L_{\odot})$$^a$ & 1.4 & 1.5 & 6.0 & 3.0 \\
$\Sigma_{\mathrm{SFR}}~(M_{\odot}\mbox{ Gyr}^{-1}\mbox{ pc}^{-2})$ & 8 & 12 & 110
& 43 \\
$P/k$ ($10^4\mbox{ K cm}^{-3}$)& 1.6 & 1.6 & 2.4 & 1.2 \\
$M_{\mathrm{CO}}~(10^5~M_{\odot})$ & 11.1 & 6.3 & 3.1 & 3.5 \\
\hline
Comparisons \\
\hline
$I_{\mathrm{HCN}} ~(\mbox{K km s}^{-1})$ & $< 0.053$ & $< 0.027$
&$\sim 0.12$ & $\sim 0.059$ \\
$I_{\mathrm{HCN},pred}~(\mbox{K km s}^{-1})$ & 0.060 & 0.090 & 0.79 & 0.310 \\
$I_{10,\mathrm{convol}}/I_{\mathrm{HCN}}$ & $>280$ & $>190$ 
& $\sim 37$ & $\sim 91$\\
$L_{\mathrm{TIR}}/L_{\mathrm{HCN}}$$^b$ & $>2000$ &
$>4200$ & $\sim 3300$ & $\sim 3300$ \\
$L_{\mathrm{TIR}}/L_{\mathrm{CO}}$$^b$ & 7.2 & 22 & 92 & 38\\
\hline
\end{tabular}
\medskip

$^a$ For conversion between intrinsic properties and quantities
derived from surface brightness, note that the $28''$ IRAM beam
subtends $1.5\times 10^4\mbox{ pc}^{2}$ at the 840 kpc distance of M33 \citep{Freedman2001}.\\
$^b$ $L_{\odot}/(\mbox{ K km s}^{-1}\mbox{ pc}^{2})$
\end{minipage}
\end{table*}

\section{Analysis and Results}
\label{analysis}
The minimal HCN emission from the galaxy was unexpected given the predicted
ratios of $I_{\mathrm{CO}}/I_{\mathrm{HCN}}$ from other studies.  For normal
galaxies, \citet{gs04} find $L_{\mathrm{CO}}/L_{\mathrm{HCN}}=25$ with $\sim
50\%$ scatter.  For GMCs in the disk in the Milky Way, \citet{hcn-MW} find
$I_{\mathrm{CO}}/I_{\mathrm{HCN}}= 40\pm 10$ but the ratio decreases for
clouds in the bulge of the galaxy to $12$.  For comparison, we find two
sources (M33GMC 91 and 26) with a dense gas ratio $>280$ and the remaining two
sources have a ratio below that of the Milky Way disk.  We conclude that the
HCN emission from M33 GMCs, relative to their CO, is a factor of 2 to 7 lower
than similar clouds in the disk of the Milky Way and other galaxies.

The lack of HCN emission becomes even more remarkable when the results are
compared with expectations based on the measured star formation rate of M33.
Previous work by \citet{wu05-hcn} suggests that the rate of ongoing star
formation is linearly related to the luminosity of HCN~($1\to 0$):
\begin{equation}
\dot{M}_{\star}~(M_{\odot}~\mbox{yr}^{-1}) 
 = 1.4\times 10^{-7} L_{\mathrm{HCN}}~(\mbox{K km s}^{-1}\mbox{ pc}^{2}).
\label{sfrhcn}
\end{equation}

We estimate the amount of HCN emission that would be predicted based
on the apparent star formation at each of our pointing locations in
M33.  Using the star formation rate estimates presented in Table
\ref{gmctable}, we use Eq.~\ref{sfrhcn} to predict the intensity of HCN
emission expected from the ongoing star formation
($I_{\mathrm{HCN},pred}$ in Table \ref{gmctable}).  We find the
predicted HCN emission is a factor of 1.1 to 6.6 higher than the
observed values or upper limits.  The discrepancy may result, in part,
from the inclusion of H$\alpha$ in the star formation indicator
vs. the infrared-only data used in Equation \ref{sfrhcn}.

\begin{figure}
\includegraphics[width=84mm]{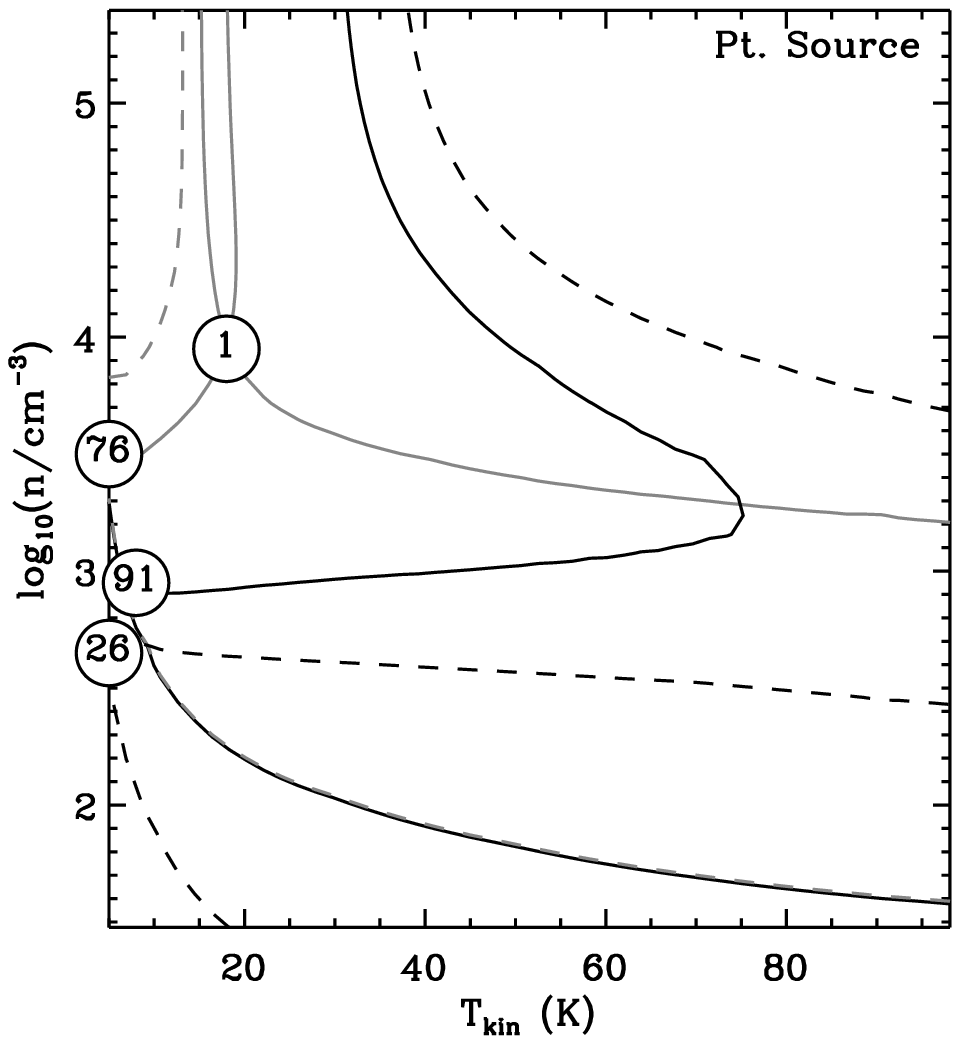}
\includegraphics[width=84mm]{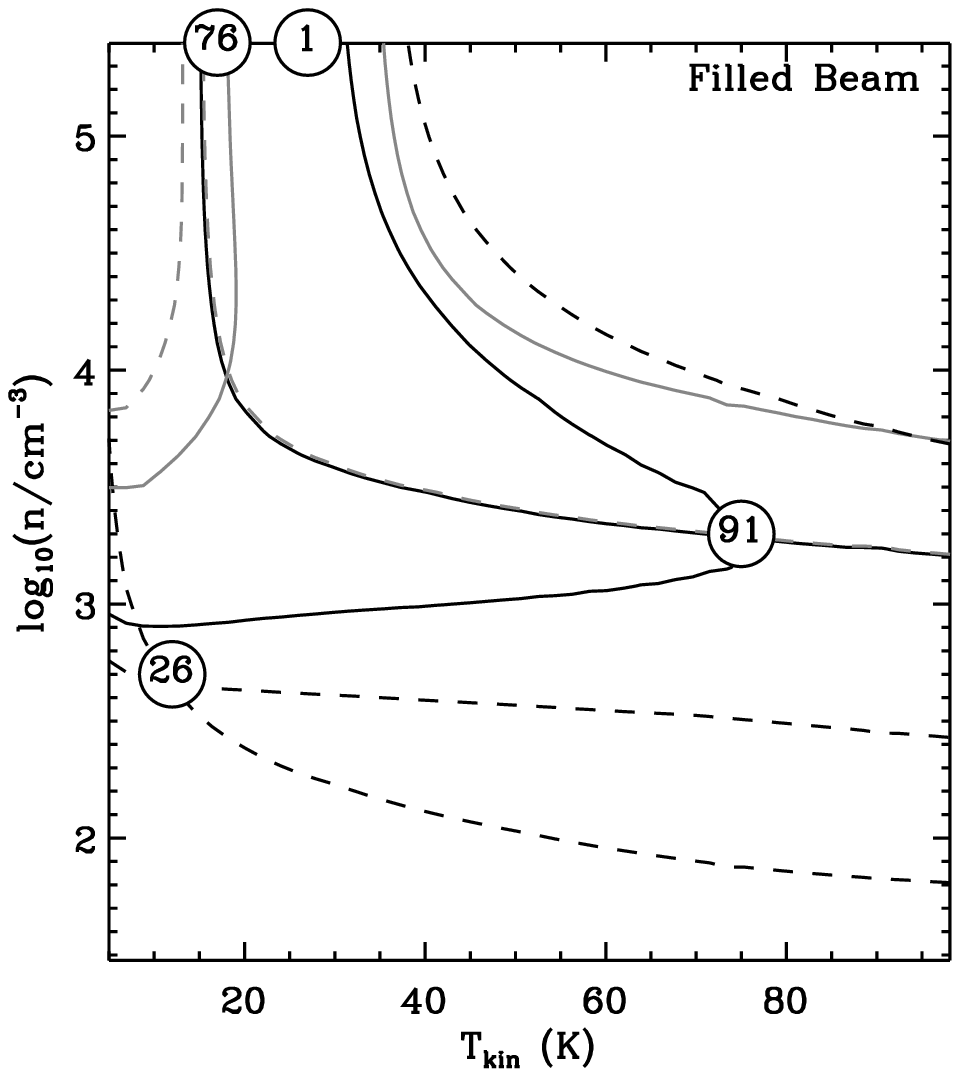}
\caption{Large Velocity Gradient analyses of the pointings in the
  observation using the RADEX code applied to the
  $^{12}\mathrm{CO}(2\to 1)$ and $(1\to 0)$ lines and the
  $^{13}\mathrm{CO}(2\to 1)$ line. The top and bottom panels indicate
  the result assuming that all the emission from the GMC targets is
  either a point source (top) or beam-filling (bottom). The lines are
  shown to indicate conditions producing a line ratio.  Intersections
  between pairs of lines show where conditions produce both line
  ratios, which are labelled with the corresponding M33GMC
  number. Points occur at the edge of the model grid if the solution
  is outside the sample box.  \label{lvg}}
\end{figure}

The $^{12}$CO$~(2\to 1)$ and $^{13}$CO$(1\to 0)$ data offer some
insight into the conditions of the molecular cloud targets.  These
line ratio data can also compared to the $I_{10,\mathrm{R07}}$
data. The difficulty in evaluating these results is the difference in
beam sizes between the $^{12}\mathrm{CO}(2\to 1)$ line ($11''$) and
the $^{12}\mathrm{CO}(1\to 0)$ line ($14''$).  If the emission fills
all beams, then the line ratio can be modelled using LVG analysis as
was done for M33 GMCs in \citet{wwt97}.  We attempt an LVG analysis
using the RADEX code \citep{radex}.  We take a typical column density
of $N_{\mathrm{CO}}\sim 10^{17.7}\mbox{ cm}^{-2}$ (assuming
$N$(CO)/$N$(H$_2=2\times 10^{-4}$) and line FWHM of $\sim 8\mbox{ km
  s}^{-1}$ representative of the sample.  We adopt
$N(^{12}$CO)/$N(^{13}$CO)=70 following \citet{wwt97} though they argue
that LVG results are relatively insensitive to this ratio.  We
calculate an LVG model for two extreme conditions: (1) the emission is
uniform within all the beams so the source-beam coupling is unity or
(2) the emission is a point source within the beam so the intensities
must be scaled by the ratio of the beam areas for comparing the lines.
We model the expected line ratios $^{12}\mathrm{CO}(2\to
1)/^{12}\mathrm{CO}(1\to 0)$ and $^{13}\mathrm{CO}(2\to
1)/^{12}\mathrm{CO}(2\to 1)$ for a grid of density ($n=10^{1.5}\to
10^{5.5}\mbox{ cm}^{-3}$) and temperature conditions ($T_K=5\mbox{
  K}\to 100 \mbox{ K}$).  We find solutions where the physical
conditions produce a consistent set of line ratios and label those
points with the M33GMC number in Figure \ref{lvg}.  The models have no
free parameters and thus are able to replicate line ratios exactly,
provided the results are sampled on the grid.  Observations of more
lines would enable a statistically robust determination of cloud
properties.  While the permitted range of conditions is broad
depending on assumptions regardings source structure, we find no
cloud that is inconsistent with all clouds being typical GMCs (i.e.,
$n\sim 10^{3}\mbox{ cm}^{-3}$ and $T \sim 15$~K).  While there is some
concern that the conditions in the clouds may not be well suited for
HCN ($1\to 0$) emission, the line ratio studies are consistent with
these objects being normal GMCs.

\section{Discussion}

The minimal HCN emission from M33 is striking, especially compared
with the active star formation from the galaxy as a whole. In
particular, M33GMCs 91 and 26 highlight the deficit of HCN emission
relative to CO when compared to Milky Way clouds throughout the
disk. We note that M33GMC 91 is the most massive cloud in M33.
M33GMCs 1 and 76 show marginal detections consistent with the outer
bounds of what is expected from Milky Way and other local galaxies.
In this section, we explore possible explanations for the high CO/HCN
ratios observed.

\begin{figure}
\includegraphics[width=84mm]{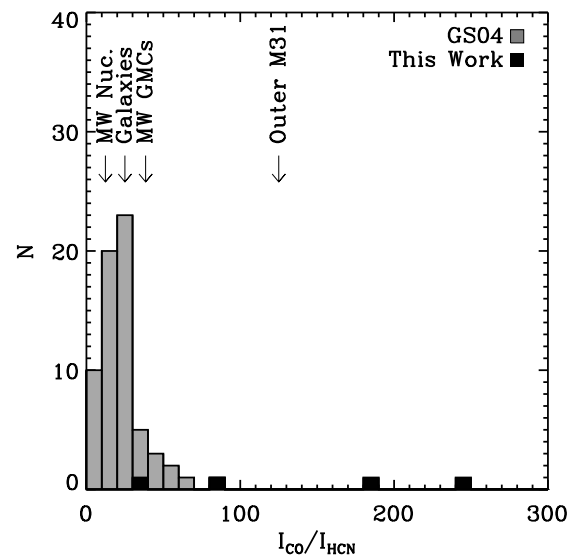}
\caption{CO-to-HCN ratios for low-$z$ galaxies \citep[][, grey]{gs04}
  and for M33 pointings (black).  The distribution shows that the HCN
  emission is anomalously weak compared to CO when viewed in the
  context of galaxies.  The highest value of the CO-to-HCN emission is
  associated with the most massive GMC in M33.\label{summ2}}
\end{figure}

The best comparable study of CO-to-HCN ratios in Local Group galaxies
is the work of \citet[][B05]{m31-hcn}, who used the IRAM 30-m
telescope to observe 16 lines of sight in M31.  Their work found a
radial decrease in $I_{\mathrm{CO}}/I_{\mathrm{HCN}}$ ranging from 30
to 125 at the outer limit (see Figure \ref{summ2}).  B05 interpreted
the radial variation as a reflection of a decrease in dense gas
content of molecular clouds at larger galactocentric radius,
consistent with the decrease seen in \citet{hcn-MW}.  However, both
M31 and the Milky Way show relatively normal star formation
efficiencies for their molecular gas content, in contrast with M33.
Further, all of the B05 measurements are in regions with
$I_{\mathrm{CO}}/I_{\mathrm{HCN}}$ and metallicity
\citep{m31-gradient} similar to the Milky Way measurements.

\subsection{Excitation Conditions}
In this work, we have focused our analysis on the $(1\to 0)$ lines of
$^{12}$CO and HCN.  Following a significant number of authors, we have
taken the CO emission to trace the ``low-density'' emission in a
molecular cloud ($n\sim 10^{2}\mbox{ cm}^{-3}$), whereas the HCN line
traces high density gas within the cloud.  The latter statement is
largely motivated by the high critical density of the transition.
However, the critical density is only a rough indicator of the
excitation conditions of the gas.  High temperatures ($T>50~\mbox{
  K}$) and significant radiative pumping can bring sharp departures
from this ``common sense.''  Without a full, multi-line analysis of
the species, a firm interpretation is impossible (\S\ref{analysis}).
However, the analysis of the physical conditions in M33 clouds suggest
that these systems are analogous to GMCs found in the Milky Way
\citep{wwt97,hcn-MW}, rather than active galactic nuclei where
substantial concerns about the interpretation of HCN have been raised.
Hence, relying on the ($1\to 0$) lines of CO and HCN as tracers of
low- and high-density gas respectively appears justified.

\subsection{Metallicity and Chemical Abundances}

The M33 system is substantially more metal poor \citep{m33-grad} with
all but the central measurement in the present study occurring at
lower metallicity than any point in the M31 study. The metallicity may
approach solar in the inner 1 kpc \citep{magrini07} which
harbours M33GMC1.  The similarities in pressure (see below) and the
presence of significant amounts of star formation in the clouds
suggests that there should be sufficient dense gas to excite HCN
emission if it is present.  The lower overall metallicity is also
related to depletion of secondary (N) elements relative to primary
elements (C,O), so HCO+ may be a more viable tracer of dense gas in
low metallicity systems.  \citet{magrini07} find a [N/O] ratio 0.3 dex
lower than solar abundances across the entire galaxy.  This depletion
may translate directly to lower HCN/H$_2$ abundances.

To evaluate the effects of lower metallicity on the relative abundance
of chemical species, we integrated a set of chemical models using the
Nahoon code \citep{nahoon}.  We considered a set of eight models
spanning a range from $\log[\mathrm{(O/H)/(O/H)_{\odot}}]=0.0$ to
$-0.4$ in uniform logarithmic steps and
$\log[\mathrm{(N/O)/(N/O)_{\odot}}]$ = 0.0 to $-0.3$ in uniform
logarithmic steps.  Solar abundances were set at the Nahoon defaults:
He/H = 0.14, C/H = $7.3\times 10^{-5}$, N/H = $2.14\times 10^{-5}$,
O/H = $1.76\times 10^{-4}$, Cl/H = $4\times 10^{-9}$, F/H =
$6.68\times 10^{-9}$, Fe/H = $3\times 10^{-9}$, Mg/H = $7\times
10^{-9}$, Na/H = $2\times 10^{-9}$, P/H = $3\times 10^{-9}$, S/H =
$8\times 10^{-8}$, Si/H = $8\times 10^{-9}$.  We scaled the solar
abundances of $\alpha$-process elements to track the O abundance (C,
Mg, S, Si) and all other elements to track N (Cl, F, P, Na). Grain
abundance was depleted by the same factor as O.  We use standard
initial conditions for chemical models, including
$n_{\mathrm{H2}}=10^4\mbox{ cm}^{-3}$ and $T=10$~K.  The models use
the 2005 OSU Astrochemistry Rate compilation distributed with the
Nahoon package.  The rates track 4423 reactions among 452 species.

We compared the abundance of several chemical species at
$t=10^7$~years (the typical lifetime of a GMC) to evaluate the effects
of depletion and low metallicity on the formation of various species,
specifically HCN and CO.  We find the CO abundance is a reasonable
value for the solar abundance model (CO/H$_2 =1.4\times 10^{-4}$) and
decreases roughly linearly with the decreasing metallicity.  Contrary
to expectations, the HCN abundance is nearly constant with decreasing
metallicity (HCN/H$_2$ = $(1.4\pm 0.1)\times 10^{-9}$), as a result
the CO/HCN ratio actually decreases by a factor of 2.5 going from
solar to M33 abundances.  These models are necessarily coarse
approximations to the actual chemistry present in clouds.  However,
they do not indicate significant increase of the CO/HCN abundance with
a decreasing N/O ratio.  The ratio HCO+/HCN is also roughly constant
with metallicity.  While not conclusive, these models are a reminder
that the abundances of individual chemical species need not track the
abundances of elements of which they are composed.  Hence, metallicity
changes do not appear to explain the dearth of HCN in M33.
\citet{hily-blant} have shown that the HCN ratio is insensitive to C/O
variations in the gas, similar to the model's N/O and O/H robustness.
A full treatment, including the effects of radiation \citep[as was
  done for CO in] []{bell-xco,glover-xco}, is needed to validate this
initial exploration.

Some authors have raised concerns about the utility of HCN as a star
formation tracer, noting that AGN activity could decrease the CO/HCN
ratio such that there would appear to be more dense gas than expected
\citep[e.g.][]{meijernik07}.  We note that this is not a concern here
because such effects invoke x-ray irradiation by an AGN and such
effects should make HCN easier rather than more difficult to detect.

Studies of HCN formation in photodissociation regions
\citep{young-owl,boger-sternberg} suggest that HCN is readily
photodissociated in translucent regions.  Their work shows that HCN
does not reach significant concentrations unless shielded by 4 to 8
magnitudes of visual extinction, depending on the local conditions.
Given that CO forms more readily ($A_V \sim 2$), translucent clouds
will show a high CO/HCN ratio.  If the subsolar metallicity in M33
predicts for lower dust-to-gas ratios, then the fraction of the volume
of the molecular cloud at high $A_V$ may be significantly reduced,
increasing the CO/HCN ratio.  Identical reasoning is forwarded to
explain the low brightness of CO in the SMC and other low metallicity
systems \citep[e.g.,][ among others]{xfac-israel, leroy-smc-xfac}.
Given the ample evidence for such an effect in low metallicity
systems, it may hold that HCN will cease to be a good tracer of star
formation at modest metallicities (e.g., $\log([\mathrm{O/H}])+12\sim
8.4$) because of dissociation effects.  Further studies of the dust
content of M33 are ongoing and should clarify the importance of
photodissociation at regulating the CO/HCN ratio on GMC scales
\citep{hermes-CO}.

\subsection{ISM Structure}

Changes in the structure of the ISM are important at setting the
CO/HCN ratio in starburst systems, where the dense gas occupies a
large fraction of the ISM \citep{gs04}.  This change reduces the
CO/HCN ratio.  It is conceivable that a similar effect is at work
here.  However, the molecular cloud population in M33 has been studied
extensively \citep{ws90,rpeb03,deepm33}, reaching the conclusion that
the individual clouds have similar macroscopic properties as those
found in the Milky Way and M31, including turbulent line widths and
average densities.  

The properties of these clouds are linked to the galactic environment via
pressure. \citet{hcn-MW} demonstrate that HCN-to-CO ratio correlates with the
disk pressure in the ISM at the galactocentric radius of the objects observed:
$I_{\mathrm{HCN}} / I_{\mathrm{CO}}\propto P^{0.19\pm 0.04}$.  Higher
pressures in the galactic environment would cause GMCs to have correspondingly
higher pressure to maintain their distinction from the remainder of the ISM.
As seen in Table \ref{gmctable}, the disk pressure values are comparable to
the Milky Way disk values, implying these GMCs do not fall on the scalings of
\citet{hcn-MW}.  Following their scaling, we predict $I_{\mathrm{CO}} /
I_{\mathrm{HCN}}\sim 30-50$ in these clouds which is only consistent with a
marginal detection from M33GMC 1.  Thus, it seems unlikely that the dense gas
content of these clouds would be significantly lower than those in the other
local group disk systems.  This is because the dense gas fraction of clouds
has a tight link between the average density and amount of turbulent driving
\citep[e.g.,][]{pn02,krum-ksslope}.

\subsection{Infrared Luminosity and Evolution}

\begin{figure*}
\begin{minipage}{126cm}
\includegraphics{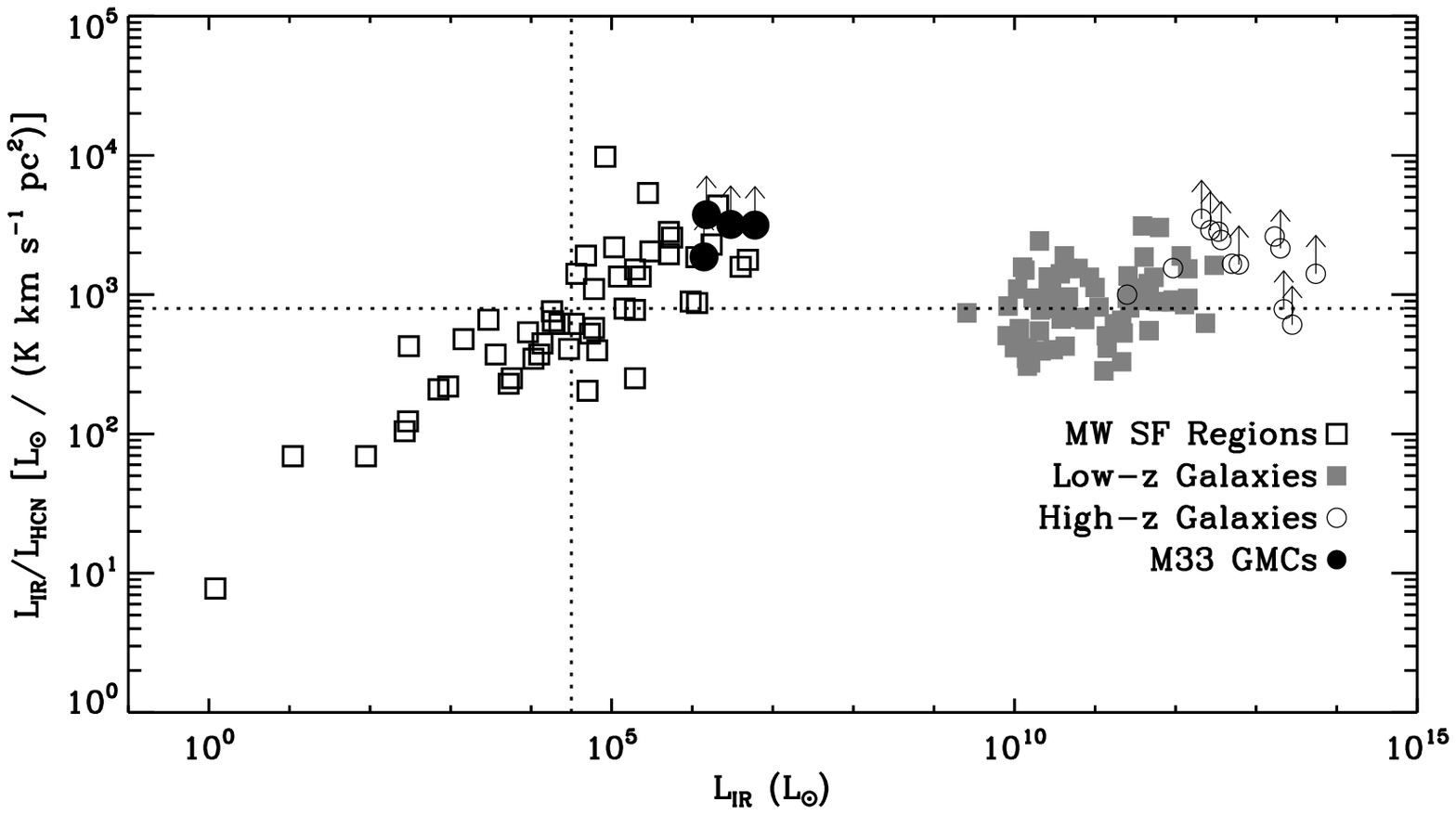}
\end{minipage}
\caption{Comparison of M33 GMCs to literature data for $L_{\mathrm{IR}}$ and
  $L_{\mathrm{HCN}}$. Milky Way star forming region data are from
  \citet{wu10}; Low-$z$ galaxies are from \citet{gs04}; high-$z$ galaxies are
  from \citet{gao07}. The horizontal line at a ratio of
  $L_{\mathrm{IR}}/L_{\mathrm{HCN}}=800$ is the mean ratio for regions with
  $L_{\mathrm{IR}}>3\times 10^{4}~L_{\odot}$ forwarded by \citet{wu10}. The
  vertical line at $L=10^{4.5}~L_{\odot}$ is attributed to the threshold below
  which the IMF is not fully sampled in the system.  The M33 GMCs are
  consistent with the upper end of the galactic population, but the regions
  are significantly above the mean ratio.\label{sumfig}}
\end{figure*}

The motivation for studying HCN in M33 comes from the intimate link
between star formation (as implied by infrared emission) and HCN
emission in galaxies.  In Figure \ref{sumfig}, we consider the
infrared and HCN emission in the context of other studies.

The targets in M33 are comparable to high luminosity star forming
regions in the Milky Way.  While they appear significantly above the
mean ratio defined by galaxies ($L_{\mathrm{IR}}/
L_{\mathrm{HCN}}=800$), the lines-of-sight appear consistent with the
upper envelope of values derived for the Milky Way regions. We note
that the observed clouds have substantial masses
\citep[$>10^5~M_{\odot}$,][]{deepm33} and their infrared luminosities
are well above the threshold above which the HCN-to-IR ratio saturates
at its constant ratio \citep[Figure \ref{sumfig},][]{wu10}.
Specifically, M33GMC 1 and 76 have bright, discrete {\sc Hii} regions
located within the IRAM beam, in addition to being strong CO emitters.
Although \citet{wu10} have argued for a constant $L_{\mathrm{IR}}/
L_{\mathrm{HCN}}$ ratio for $L_{\mathrm{IR}}>10^{4.5}~L_{\odot}$, the
addition of the M33 points to the upper envelope of the trend suggest
a continuation to higher ratios at larger luminosities.  Hence, the
remarkably consistent values for the IR-to-HCN ratio seen in galaxies
may actually derive from consistent sampling from a population with a
changing ratio.

The relationship between star formation and dense molecular gas
emission is not universal.  Indeed, \citet{gao07} found that four high
redshift galaxies ($z=1.0 \to 2.8$) show
$L_{\mathrm{HCN}}/\dot{M}_{\star}$ that is a factor of 2 lower than
the result established in the local population and several
non-detections of HCN suggest a cosmological evolution of the star
formation efficiency of dense gas.  While the GMCs in M33 do not have
star formation efficiencies as high as observed in some high redshift
systems, the results are consistent with the star formation efficiency
of HCN-emitting gas being comparable to or larger than that observed
in high redshift systems.

This study observes individual GMCs whereas the other studies focus on
galaxies \citep{gs04,gao07} or individual star forming regions
\citep{wu05-hcn}.  Thus, the evolution of GMCs via cloud evolution and
star formation processes may cause scatter in the interpretation of
star formation efficiency.  We have selected GMCs such that
$L_{\mathrm{IR}}/L_{\mathrm{CO}}$ spans a large range.  While not
conclusive, it is suggestive that the two marginal detections showing
the lowest $I_{\mathrm{CO}} / I_{\mathrm{HCN}}$ ratios are also the
most infrared luminous.  The lack of dense gas in M33GMCs 91 and 26
may indicate these are younger, less evolved systems.  The
factor-of-ten range in star formation rates derived from the IR +
H$\alpha$ emission suggests that variations in evolutionary stages may
indeed be responsible for the low levels of HCN emission seen from
M33GMCs 91 and 26.  Thus, a possible explanation for the highest
values of CO-to-HCN may be that these clouds are young and have not
formed significant fractions of high density gas.

\section{Conclusions}

In summary, we have searched for and failed to find significant HCN emission
from GMCs in M33.  The HCN observations are sufficiently sensitive that the
HCN should be easily detected ($>20\sigma_{rms}$) if the emission were present
at the levels expected based off either CO or star formation observations in
our Galaxy and others \citep{hcn-MW, gs04, wu05-hcn}.  Two of the four lines
of sight that we observed show a marginal detection of HCN emission and these
two targets are bright infrared sources in the galaxy.  Using na\"ive chemical
models, we explored the effects of the low metallicity and N/O depletion in
M33 on the CO/HCN ratio.  Despite depletion of N/O, the CO/HCN ratio is
predicted to be lower in M33 than in the Milky Way.  Since HCN only forms at
high extinction relative to C/O, the low metallicity of the galaxy may result
in reduced dust shielding and thus more molecular gas exposed to dissociating
radiation.  Even though the ratio of $L_{\mathrm{IR}}/L_{\mathrm{HCN}}$ is a
factor of six larger than the mean for galaxies, it is nonetheless consistent
with comparable systems in the Milky Way.  The M33 targets appear most
distinct in their CO-to-HCN ratios rather than in their infrared luminosities.
This suggests that the low levels of HCN emission likely reflect a smaller
fraction of dense, shielded gas in these clouds rather than a systematic
effect.  Having considered chemical and structural variations, the observed
discrepancies seem most likely to be the result of either reduced extinction
or cloud evolution effects.  A larger sample of GMC-based observations and a
clearer understanding of the dust properties in M33 are needed to clarify
these results.

ER and JP are supported by an NSF AAP Fellowship (AST-0502605). ER is
further supported by a Discovery Grant from NSERC of Canada. YG's
research is partially supported by China NSF Innovation Team
(\#10621303 ), Distinguished Young Scholars (\#10425313), 973 of the
Ministry of Science \& Technology and Chinese Academy of Sciences'
Hundred Talent Program.  We are grateful for the comments of an
anonymous referee, in particular regarding the importance of
UV dissociation at regulating the CO/HCN abundance.  


\end{document}